\documentstyle[11pt]{article}
\newcommand{\be}{\begin{eqnarray}}

\newcommand{\ee}{\end{eqnarray}}

\title{
	\begin{flushright}
	{\normalsize TPI--MINN--94--40/T \\
	NUC--MINN--94--6/T \\
	HEP--MINN--94--1321 \\
        DOE/ER/40561-175-INT94-00-78\\
	January 1995 \\}
	\end{flushright}
\bf The Gluon Propagator in non--Abelian Weizs\"{a}cker--Williams Fields
}
\author{
	Alejandro Ayala, Jamal Jalilian-Marian and Larry McLerran \\
	{\small\it School of Physics and Astronomy,
	University of Minnesota, Minneapolis, MN 55455} \\
	Raju Venugopalan \\
	{\small\it Institute for Nuclear Theory,
	University of Washington,
	Seattle, WA 98195 } \\
 }

\date{}

\begin{document}

\maketitle

\begin{center}
{\bf Abstract}\\
\end{center}

We carefully compute the gluon propagator in the background of a non--Abelian
Weizs\"{a}cker--Williams field.   This background field is generated by the
valence quarks in very large nuclei.  We find contact terms in the small
fluctuation equations of motion which induce corrections to a previously
incorrect result for the gluon propagator in such a background field. The well
known problem of the Hermiticity of certain operators in Light Cone gauge is
resolved for the Weizs\"{a}cker--Williams background field. This is achieved
by working  in a gauge where singular terms in the equations of motion are
absent and then gauge transforming the small fluctuation fields to Light Cone
gauge.

\vfill \eject

\section{Introduction}

In a recent work, two of the authors have argued that parton distribution
functions may be computed for very large nuclei at small x~\cite{mv}. They
found that the valence quarks in these nuclei are the sources of non--Abelian
Weizs\"{a}cker--Williams fields. This work is similar in spirit to that of A.
Mueller who considers the distribution functions generated by heavy quark
charmonium--like bound states~\cite{al}. In reference~\cite{mv} the general
form of these fields was found.  It was argued that to compute the distribution
functions, one should compute the fields associated with these valence quarks
and then color average over these fields treating the valence quarks as
classical charges.  The weight for this averaging was shown to be
\be
	\exp\left( - {1 \over {2\mu^2}}\int d^2x_t \rho^2(x_t) \right)
\ee
where $\rho(x_t)$ is the distribution of transverse valence quark color charge
and $\mu^2$ is a parameter corresponding to the valence quark color charge per
unit area.

The current for the valence quarks was found to be
\be
	J^{\mu}_a (x) = \delta^{\mu +} \delta(x^-) \rho_a(x_t,x^+)\,.
\ee
Here the Light Cone variables are
\be
	y^{\pm} = (y^0 \pm y^3)/\sqrt{2}\,.
\ee
The approximation which leads to this form of the current is that $x <<
A^{-1/3}$ so that the small x partons see a nucleus which is essentially
Lorentz contracted to a thin sheet.  This is the origin of the factor $\delta
(x^-)$ above. Further, since the small x partons are weakly coupled to the
source, the source is ``static" and only the $+$ component of the current is
important.

The charge distribution in transverse coordinates is essentially random with a
Gaussian weight as described above. This is true so long as we look at
transverse resolution scales which are large enough that there are many valence
quarks per unit area, and small enough that it is much less than the
confinement scale of $1$ fm.  In momentum space, this requires that we only
consider transverse momenta which are in the range
\be
	\Lambda_{QCD}  << k_t << \mu \sim A^{1/6}\,.
\ee
The $x^+$ (Light Cone time) dependence of the charge density is a consequence
of the extended current conservation law
\be
	D_\mu J^\mu = 0.
\ee

The current may be taken to be time independent if we fix the residual gauge
freedom by the constraint that $A^- (x^- = 0) = 0$. This point will be
discussed further later.

Under these conditions, it is easy to guess the form of the
Weizs\"{a}cker--Williams field.  The charge distribution is confined to an
infinitesimally thin sheet.  Therefore, one expects that the fields on either
side of the sheet might be a gauge transform of vacuum.  We deduce therefore
that~\cite{mv}
\be
	A^+ & = & 0 \nonumber \\
	A^- & = & 0 \nonumber \\
	A^i & = & \theta (x^-) \alpha^i(x_t)\label{eq:trans}
\ee
where
\be
	\alpha^j (x_t) = {i \over g} U(x_t) \nabla^j U^\dagger (x_t).
\ee

The field $\alpha$ is a gauge transform of the vacuum.  It generates zero
magnetic field.  There is a transverse electric field which is confined to the
sheet $x^- = 0$.  Gauss's law shows that this field solves
\be
	\nabla \cdot \alpha = g \rho \,.
\ee
One can verify that the Light Cone Hamiltonian satisfies
\be
	P^- = 0
\ee
so that these classical fields have vanishing Light Cone energy.

These fields are gauge transforms of vacuum fields with a discontinuity which
is adjusted so as to reproduce the correct valence charge distribution. Because
it is a pure gauge transform of the vacuum on either side of the
infinitesimally thin sheet, we expect that the propagators are simple. In a
previous paper, two of the authors argued that the fields which generate the
propagator are just gauge transforms of the ordinary small fluctuation plane
wave fields~\cite{mv2}.  This argument is correct for matter fields such as
scalars and fermions, but as we shall see is not quite correct for gluon
fields.

The problem with the gluon fields is that there can be contact terms in the
gluon small fluctuation equations of motion associated with the thin sheet of
valence charge.  These contact terms can generate contributions to the gluon
field which are constant or grow linearly at large distances away from the
sheet.  The linearly growing terms will give rise to a constant electric field.
In any case, neither of these terms were allowed for in the previous analysis
which claimed to have determined the gluon propagator.

The terms which were ignored in the previous analysis are very singular. A
direct consequence of their bad behavior at infinity is the non--self--adjoint
character of the Light Cone gauge equations of motion.

The present work is organized as follows: in Section 2 we will see that the
above mentioned terms are a consequence of the small fluctuation equations of
motion in Light Cone gauge ($A^+ = 0$). To resolve the ambiguities inherently
associated with the singular behavior of the small fluctuation fields, we
choose to work in $A^- = 0$ gauge. In Section 3, we show that this gauge is
non--singular and that the propagator can be computed directly.  It turns out
that the propagator in this gauge is precisely that previously computed by
McLerran and Venugopalan \cite{mv2}. The propagator in Light Cone gauge is
derived from  the propagator in the non--singular gauge.
Details of this derivation are the subject of Section 4. In section 5, we
discuss the
regularization prescriptions which enable us to generate a properly defined
propagator in Light Cone gauge. We write down an explicit expression for this
propagator. The propagator will be used in a sequel to this work to calculate
the gluon distribution function to next order in $\alpha_{s}$.

In Section 6, we summarize our results.

\section{The small fluctuation equations in Light Cone gauge}

The conventional approach to computing a Green's function is to write down the
equations for the small fluctuation field in a fixed gauge.  Denoting the small
fluctuation field by $\delta \phi$, the generic small fluctuation equation is
\be
	G^{-1} \delta \phi_\lambda = \lambda \delta \phi_\lambda \label{eq:def}
\,,
\ee
and the propagator is
\be
	G(x,y) = \sum {{\delta \phi_\lambda (x)
\delta \phi^\dagger_\lambda (y) } \over \lambda}\,.
\label{eq:cstrt}
\ee

To derive the equations for the small fluctuation field in Light Cone gauge
($A^+=0$) we first look at equation~(\ref{eq:def}) with $\lambda = 0$.
Expanding around our background field, we find that the equations for the gluon
field become
\be
	\left( D(A)^2 g^{\mu \nu} - D^\mu (A) D^\nu (A) \right)
\delta A_\nu - 2 F^{\mu \nu} \delta A_\nu = 0.
\ee
In this equation, $A$ is the background field which according to
equation~(\ref{eq:trans}) is non--vanishing only for its transverse components.
$D^{\mu}(A)$ is the covariant derivative with $A$ as the gauge field (notice
that $D^{\pm} = \partial^{\pm}$).  We have assumed the gauge condition $\delta
A^- (x^- = 0) = 0$ so that we can ignore any time evolution of the source of
valence quark charge.  If this gauge condition were not chosen, then the
extended current conservation condition would require that the source evolve in
time, and this would generate a correction to the above equation. This
background field generates no magnetic field, but it does generate a non--zero
electric field, $F^{+i}$
\be
	F^{i+} = \delta(x^-)\alpha^i(x_t).
\ee

On either side of $x^- = 0$, the background field is a gauge transform of the
vacuum.  We might therefore think that the solutions to the above equation, in
the two different regions of $x^-$, are simply different gauge transforms of
the plane wave small fluctuation field one would get in zero background field.
Further, continuity at the origin could be assured by a matching condition.
This is precisely what was done in the previous work of McLerran and
Venugopalan \cite{mv2}.

Unfortunately, this procedure will not work here due to the presence of the
background electric field term.  Before seeing how this works, let us first
express the small fluctuation equations in a slightly more transparent form
(hereafter latin indices refer to transverse variables).
\be
	D^2(A) \delta A^\mu - D(A)^\mu D(A) \cdot \delta A =
	   -2\delta^{\mu +}\delta (x^-) \alpha^i \times \delta A_i
	   -2\delta^{\mu j} \delta(x^-) \alpha^j \times \delta A^- .\nonumber
\ee
The terms in the right--hand side of the above equation are called contact
terms,
they appear when the derivative $\partial/\partial x^-$ acts on the classical
field which is proportional to $\theta (x^-)$.
Notice that the last term vanishes in the gauge where $\delta A^- \mid_{x^- =
0} = 0$.

The small fluctuation equation  for the plus component of the above equations
becomes
\be
	\partial^+ (\partial^+ \delta A^- - D(A)^i \delta A_i )
	    = -2 \delta (x^-) \alpha^i \times \delta A_i.
\ee
This equation requires that
\be
	   \partial^+ \delta A^- - D(A)^i \delta A_i
	    = -2 \theta (x^-) \alpha^i \times \delta A_i\mid_{x^- = 0}
	       +\,\,h(x^+,x_t)
\ee
where $h$ is some as yet undetermined function.  We see immediately that this
implies that the fields have singular behavior at infinity. We might remove the
singular behavior at say $x^- \rightarrow -\infty$ but due to  the step
function, cannot remove it as $x^- \rightarrow +\infty$. In fact, the fields
will behave like $x^-$ as $x^-$ goes to infinity. This singular behavior is
sufficient to guarantee that the equations of motion are not self--adjoint on
this space of functions. Further, the small fluctuation fields are
discontinuous at the origin ($x^-=0$) and matching the fields there is more
difficult. The singular behavior at infinity and the discontinuity of the
fields at the origin were absent from the previous analysis of McLerran and
Venugopalan~\cite{mv2}. This was because they neglected to include the contact
term at the origin which (as we have seen) is present in the equation of motion
in Light Cone gauge.

Due to this lack of self--adjointness on this space of functions, we have found
no consistent way of directly integrating the equations of motion in Light Cone
gauge. The most obvious problem is that the Green's function will not solve the
constraint imposed by symmetry and by the Hermiticity of the underlying fields.
Without a better way of defining the equations of motion, we are at a quandary
and do not know how to properly construct the Green's function in Light Cone
gauge. This problem with Light Cone gauge is well known in the relevant
literature and various schemes have been adopted to circumvent it~\cite{kallo}.

However, and this is the core of the present analysis, we are compelled to
compute the Green's function in Light cone gauge. There are several reasons
for this:

It is well known that only in Light Cone gauge does the parton model have the
most
direct and manifest physical realization~\cite{AHM}. Only in Light Cone gauge
it is possible to construct a simple and intuitive Fock space basis on which
our formalism relies heavily. Light Cone gauge (like any other axial gauge) is
free of ghosts and negative-norm gauge boson states which makes the theory
simpler~\cite{ptQCD}.

In addition, there are more advantages in choosing the $A^{+} =0$ gauge. It can
be shown that in this gauge $J^{+}$ is the only large component of the
fermionic current. The transverse and minus components of the current are
proportional to
${1 \over P^{+}}$ and are therefore small~\cite{mv}. It is much easier
to eliminate the dependent fermionic degrees of freedom in this gauge since
we don't need to worry about inverting  covariant derivatives. Finally,
the gluon distribution function is most simply related to physical quantities
measured in deep inelastic scattering experiments only in
$A^{+}=0$~\cite{brodsky}.

We will demonstrate in the following sections that the Green's functions in
Light Cone gauge can be obtained by the following procedure: i) first, solve
the small fluctuation equations in an alternative gauge ($A^-=0$) and then ii)
gauge transform the Green's functions thus obtained to Light Cone gauge. In the
following section, we will show that the Green's function may be constructed in
$A^-=0$ gauge and that the equations of motion for the Green's function are
self--adjoint.  In subsequent sections we discuss the mechanics of how the
Green's function may then be gauge transformed to Light Cone gauge.  When this
transformation is carried through, there is in general a singularity of the
transformation function. This singularity is regulated using the well known
Leibbrandt--Mandelstam prescription~\cite{lm}. We find that the Green's
function in Light Cone gauge does not satisfy the constraints which arise from
self--adjointness of the gluon field. This lack of self--adjointness is not
irretrievably damaging and has been dealt with in the past for Green's
functions of gluon fields in Light Cone gauge~\cite{bass}.
Presumably the precise nature of the singularity is not important for gauge
invariant quantities.  When one chooses a prescription for regulating the
singularity, if performed consistently in all quantities, it will give gauge
invariant results for physical quantities.  We are not aware that there is a
proof of this, but believe it to be so.

\section{The small fluctuations propagator in $A^- = 0$ gauge}

What is the virtue of writing down the equations of motion in $A^- = 0$ gauge?
In the small fluctuation equations for the transverse fields, the source term
vanishes (the equation reduces to the sourceless Klein--Gordon field equation)
\be
\left[-2\partial^+\partial^- + D_t^2(A)\right]\delta A_i = 0 \, .
\ee
This  equation is obtained  by appropriately fixing the residual gauge freedom
in $A^- =0$ gauge. One of the other equations is also sourceless, and is the
Gauss law constraint which must be imposed on initial field configurations.
This constraint commutes with the equations of motion.  It is
\be
	\partial^- \left(\partial^- \delta A^+ - D_i(A) \delta A_i\right) = 0.
\ee

We must now check that the remaining equation of motion is properly satisfied.
If we use the Gauss law constraint that $D^\mu (A) \delta A_\mu = 0$, this
equation is
\be
	\left[-2\partial^+\partial^- + \vec{D}^2(A)\right]\delta A^+ = -2 \delta
(x^-) \alpha^i \times \delta A^i.
\ee
Now from the Gauss law constraint,
\be
       \delta A^+ = {1 \over \partial^-} \vec{D}(A) \cdot \delta \vec{A}.
\ee
Although $\delta A^i$ is continuous at $x^- = 0$, the spatial covariant
derivative operator is not.  The discontinuity gives precisely the source term
on the right hand side of the equation, and the equations are consistent.

We can now solve the above equations of motion in the following way. We assume
that the transverse field on the side where the background field vanishes is a
simple plane wave.  On the other side of the boundary where the background
field is a gauge transform of the vacuum, we take the small fluctuation field
to be a linear superposition of plane waves. The linear superposition is done
in a manner to ensure that the fields are continuous across $x^- = 0$.  This is
possible because the equations above have no delta function singularity at $x^-
= 0$.

The expression for the transverse components of the fluctuation field in
$A^{-}=0$ gauge is
\begin{eqnarray}
\tilde{A}_i^{\alpha\beta}&=& e^{ipx}\eta_i\bigg\{\theta(-x^-)
\tau^{\alpha\beta}+ \theta(x^-) \int {{d^2 q_t} \over {(2\pi)^2} }e^{iq_t x_t}
\nonumber \\ &\times & \exp{\left[-i{{2p_tq_t+q_t^2}\over 2p^-}x^-\right]}\int
d^2 z_t e^{-iq_t z_t}\left(U(x_t) U^{\dagger}(z_t)\tau U(z_t) U^\dagger
(x_t)\right)^{\alpha\beta}\bigg\} \, .\nonumber \\
\label{tvec}
\end{eqnarray}
This solution is continuous at $x^{-}=0$ and was fully discussed in an earlier
paper by McLerran and Venugopalan~\cite{mv2}.

The above expression can be substituted in equation~(\ref{eq:cstrt}) to obtain
the gluon propagator in $A^-=0$ gauge. There are four separate pieces for
$G_{ij}^{\alpha \beta ;\alpha^\prime \beta^\prime} (x,y)$ which depend upon the
signs of $x^-$ and $y^-$. The overall expression for $G$ is
\be
\!\!\!G_{ij}^{\alpha \beta;\alpha^\prime \beta^\prime}\! (x,y)
\!\!\!\!\!& = &\!\!\!\!\! -\delta_{ij}\int {{d^4p} \over {(2\pi)^4}}
{{e^{ip(x-y)}} \over {p^2-i\epsilon}}
\Bigg\{ \theta(-x^-) \theta(-y^-) \tau_a^{\alpha \beta} \tau_a^{\alpha^\prime
\beta^\prime} \nonumber \\
\!\!\!\!\!& + &\!\!\!\!\! \theta(x^-) \theta(y^-) F_{a}^{\alpha \beta}(x_t)
F_{a}^{\alpha^\prime  \beta^\prime}(y_t) +
 \int {{d^2q_t} \over {(2\pi)^2}} d^2 z_t \nonumber \\
\!\!\!\!\!& \times &\!\!\!\!\!
\bigg[\theta(x^-) \theta(-y^-) e^{i(p_t-q_t)(z_t-x_t)}
e^{-i(q_t^2 - p_t^2)x^-/2p^-} F_{a}^{\alpha \beta}(x_t)
F_{a}^{\alpha^\prime  \beta^\prime}(z_t) \nonumber \\
\!\!\!\!\!& + &\!\!\!\!\! \theta(-x^-) \theta(y^-) e^{i(p_t-q_t)(y_t-z_t)}
e^{i(q_t^2 - p_t^2)y^-/2p^-} F_{a}^{\alpha \beta}(z_t)
F_{a}^{\alpha^\prime  \beta^\prime}(y_t) \bigg] \Bigg\} \label{OldGreen}
\ee
where $F_{a}^{\alpha
\beta}(x_{t}) = \bigg( U(x_{t})\tau_{a}U^{\dagger}(x_{t})\bigg)^{\alpha\beta}$.
It can be easily checked that this Green's function satisfies
\be
\left[-2\partial^+\partial^- + D_t^2(A)\right]_{ik}\,G_{kj}(x,y)\,=
\,\delta_{ij}\,\delta ^{(4)} (x-y)\nonumber
\ee
as it must. This construction of the Green's function was previously performed
by McLerran
and Venugopalan in reference~\cite{mv2}.  However, they incorrectly interpreted
it
to be the result for $A^+ = 0$ gauge, while we have shown here that it applies
to $A^- = 0$ gauge only.

For the pieces of the propagator which involve the component $\delta A^+$ we
have
\be
	<\delta A^i \delta A^+> = <\delta A^i {1 \over \partial^-} \vec{D}(A)
         \cdot \vec{\delta A}>
\ee
and
\be
	<\delta A^+ \delta A^+> = < {1 \over \partial^-} \vec{D}(A) \cdot
         \vec{\delta A}
\,\,{1 \over \partial^-} \vec{D}(A) \cdot \vec{\delta A} > - \,i\delta G^{++}
\ee
where in our convention $<A_{\mu}A_{\nu}>=\,-i\,G_{\mu \nu}$ and
$<A_{\mu}A_{\nu}>$ represents the correlation of fields in the path
integral sense.

\section{On deriving the $A^+ = 0$ propagator from the $A^- = 0$
propagator}

In the previous section we derived an explicit expression for the small
fluctuation Green's function in $A^-=0$ gauge. In this section, we discuss
general arguments which allow us to compute expectation values of operators in
one gauge in terms of the expectation values of operators in another gauge. The
formal arguments will be illustrated by the specific example of the free gluon
propagator. We are of course interested in these arguments because we want to
find  the gauge transform which will relate the Green's function in Light Cone
gauge to the Green's function in $A^-=0$ gauge.

Let $O(A)$ be some function of the field operators evaluated in Light Cone
gauge, $A^+ = 0$.  We will here consider QED and later generalize to the case
at hand. We have to evaluate the expectation value
\be
	<O(A)> = {{\int [dA] ~\delta(A^+)~ e^{iS} O(A)} \over
{\int [dA]~ \delta(A^+)~ e^{iS}}}. \label{eq:der1}
\ee
We now introduce the identity in the form
\be
	1 = {{\int [d\Lambda] ~\delta(A^- + \partial^-\Lambda)}
\over {\int [d\Lambda]~ \delta(A^{-}+\partial^-\Lambda)}}.
\ee
Inserting this into the equation for $O(A)$, we find that
\be
	<O(A)> = {{\int [dA][d\Lambda]~ \delta(A^- +\partial^-\Lambda)~
 \delta(A^+)~ e^{iS} O(A)} \over
{\int [dA][d\Lambda]~ \delta(A^- +\partial^-\Lambda)~ \delta(A^+)~ e^{iS}}}.
\ee
We now perform the gauge transformation
\be
	\tilde{A}^{\mu } = A^\mu + \partial^\mu \Lambda .
\ee
Since both the measure and the action are invariant under the gauge
transformation, we see that
\be
	<O(A)> = {{\int [d\tilde{A}][d\Lambda]~ \delta(\tilde{A}^-)
{}~\delta(\tilde{A}^+ - \partial^+\Lambda)~ e^{iS}
O(\tilde{A} - \partial \Lambda)}
\over
{\int [d\tilde{A}][d\Lambda] ~\delta(\tilde{A}^- )
\delta(\tilde{A}^+ - \partial^+\Lambda) e^{iS}}}.
\ee
Now doing the integral over $\Lambda$, we get
\be
	<O(A)> = {{\int [d\tilde{A}] ~\delta(\tilde{A}^-)~  e^{iS}
O(\tilde{A} - \partial \left({{\tilde{A}^+} \over {\partial^+}}
\right))}
\over {\int [d\tilde{A}] \delta(\tilde{A}^- )e^{iS}}}.\label{eq:GC}
\ee

Before proceeding any further, we will be more explicit and illustrate the
above approach by an example: we show that the free propagator in the absence
of a background field in $A^+ = 0$ gauge can be obtained as a gauge
transformation of the same quantity in $A^- = 0$ gauge.

Let us first review the construction of the propagator in $\tilde A^- = 0$
gauge. Henceforth all fields in this gauge will be denoted by a $\tilde A$ to
distinguish them from the corresponding fields in Light Cone gauge. The action
for the vector field is
\be
	S = { 1 \over 2} (\partial^-\delta \tilde A^+ - \nabla \cdot \delta \tilde
A)^2
	    +{ 1 \over 2} \delta \tilde A^{i} \delta_{ij} \Box \delta \tilde A^{j} \,,
\ee
where the operator $\Box$ is the D'Alambertian in Light Cone coordinates. From
the above
expression for the action, we distinguish two classes of small
fluctuation fields that satisfy the equations of motion. We can see this from
the following
argument. To find the solution that minimizes the action let us perform the
shift
\be
\delta \tilde A^+ \rightarrow \delta {\bar{A}}^+ + {1\over
\partial^-}{\nabla\cdot \delta \tilde
A}\,,
\ee
to obtain a purely quadratic action in $\delta {\bar{A}}^+$ and
$\delta \tilde A_i$.  The minimum thus corresponds to the fields that make the
action
vanish. We label the fields as follows: \\
type I fields satisfy
\be
   \partial^-\delta \tilde A^+ - \nabla \cdot \delta \tilde A & = & 0 \nonumber
\\
   \Box \delta \tilde A^i & = & 0 \label{eq:type I}
\ee
\noindent whereas type II fields satisfy
\be
   \delta \tilde A^i & = & 0  \nonumber \\
   \partial^- \delta \tilde A^+ & = & 0 \label{eq:type II}.
\ee
The contributions of different components to the free propagator
are then
\be
	<\delta \tilde A^i \delta \tilde A^j> & = & i{\delta^{ij}\over k^2} \nonumber
\\
	<\delta \tilde A^i \delta \tilde A^+> & = & i{k^i \over {k^-k^2}} \nonumber \\
	<\delta \tilde A^+ \delta \tilde A^+> & = & i\left({k_t^2 \over {(k^-)^2
k^2}}-{1\over (k^-)^2} \right)\,\,.
\ee
The second term in the parenthesis above corresponds to the contribution of the
expectation value $<\delta{\tilde A}^{+}_{II}\delta{\tilde A}^{+}_{II}>$ to
$<\delta \tilde A^+ \delta \tilde A^+>$. Notice that the contribution from
terms like $<\delta{\tilde A}^{+}_{I}\delta{\tilde A}^{+}_{II}>$ vanish. If
we now combine all the above terms together we obtain the familiar expression
for the gluon propagator
\be
	G^{\mu \nu } = - {1 \over k^2}
\left( g^{\mu \nu} -{{n^{\ast\mu}k^\nu + k^\mu n^{\ast\nu}}
\over {k \cdot n^{\ast}}} \right)\,,
\ee
where $n^{\ast\mu}=\delta^{\mu +}$.

Now perform the gauge transformation
\be
	 \delta A^\mu = \delta \tilde A^\mu + \partial^\mu \Lambda \, .
\ee
Since we are gauge transforming to Light Cone gauge, we require
\be
	\partial^+ \Lambda = -\delta \tilde A^+ \,.
\ee
We therefore obtain
\be
	\delta A^+ & = & 0 \nonumber \\
	\delta A^- & = & -{\partial^- \over \partial^+} \delta \tilde A^+ \nonumber \\
	\delta A^i & = & \delta \tilde A^i - {\partial^i \over \partial^+} \delta
\tilde A^+
        \label{eq:Gaugetrans}.
\ee
We now use equation~(\ref{eq:GC}) and the previous expression for the
propagator in $\tilde A^- =0$ gauge to derive
\be
	G^{\mu \nu } = -{1 \over k^2}
\left( g^{\mu \nu} -{{n^{\mu}k^\nu + k^\mu n^{\nu}}
\over {k \cdot n}} \right)
\ee
where $n^\mu=\delta^{\mu -}$. This result is the well known expression for the
free gluon propagator in Light Cone gauge.

For the case at hand, small fluctuations around a background field, the
analysis goes through almost exactly as before. There are two essential
differences.  The first is that for the type of background field we consider,
the gauge transformation property of the transverse field is
\be
	\delta A^i = \delta \tilde{A}^i + D^i(A) \Lambda.
\ee
One should note that the particular form of the gauge transform above is valid
only because we are interested in small fluctuations.  We are throwing away
terms which correspond to higher powers of the small fluctuation field. For the
plus and minus components, the transformation involves ordinary derivatives.
We see that the analysis above goes through in the same way it did for QED.

The other difference from the QED case is that the structure of the small
fluctuations analysis is more complicated.  There are modes (labeled
type I) which involve both $\delta \tilde A^i$ and $\delta \tilde A^+$ as
before but these
modes now couple to the background field (see equation~(\ref{eq:Gaugetrans})
with $\partial^i$
replaced by $D^i(A)$). There are also modes (type II) which
involve only $\delta \tilde A^+$ and therefore do not couple to the background
 field since
$\delta \tilde A^i = 0$. The contribution of the type I modes to the propagator
arises from the distortion of the small fluctuation fields by the background
field. The contribution of the type II modes to the gluon propagator in
momentum space is
\be
	\delta G^{++}_{(II)} &  = & {1 \over {(k^-)^2}} \nonumber \\
		           & = & {1 \over k^2} \left(
{{k_t^2} \over {(k^-)^2}} - 2{k^+ \over k^-} \right)
\ee
The second of the two expressions in the above equations is algebraically
equivalent to the first but corresponds to a different way of regulating the
singular behavior of the type II modes at $k^- = 0$. In the next section, we
will see that the latter choice is the one aptly suited to regularization.

\section{Regularization prescriptions and results}

We shall now use the method described in the previous section to derive the
gluon propagator in Light Cone gauge from the propagator in $\tilde A^-=0$
gauge (equation~(\ref{OldGreen})). Note that the gluon propagator here is the
gluon propagator in the background field. It will shortly become clear that we
need to choose an appropriate regularization prescription to obtain the final
expression for the propagator.

If we have a fluctuation field in $\tilde A^- = 0 $ gauge , the fields in $A^+
= 0$ gauge are given in terms of the tilde fields as
\be
	\delta A^+ & = & 0\\
	\delta A^- & = & - {\partial^{-} \over {\partial^+}} \delta
\tilde{A}^+\\
	\delta A^i & = & \delta \tilde{A}^{i} - D^i(A)
{1 \over {\partial^+}} \delta \tilde{A}^+.
\label{eq:GT}
\ee
To derive these equations, we note that a small gauge transformation with
a fixed background field $A$, where $A$ has only transverse components, is
\be
	\delta A^+ & \rightarrow & \delta A^+ + \partial^+ \Lambda \\
	\delta A^- & \rightarrow & \delta A^- + \partial^- \Lambda \\
	\delta A^i & \rightarrow & \delta A^i + D(A)^i \Lambda.
\ee
We have used the fact that in $A^- = 0$ gauge
\be
	\partial^- \delta \tilde{A}^+ = D(A)_i \delta \tilde{A}_i.
\ee

The transverse fluctuation field in Light Cone gauge is then
equation~(\ref{eq:GT}) which we rewrite in matrix component notation as
\be
\delta A_{i}^{\alpha \beta}(x)=
\bigg[\delta _{ik}-D_{i}\frac{1}{\partial _{-}\partial _{+}}
D_{k}\bigg]^{\alpha \gamma}\,\,\,\delta \tilde{A}^{\gamma \beta}_{k}(x).
\ee
To see how the covariant derivative acts on the field, notice that
$D_{k}\delta \tilde{A}_{k}(x)=
\partial _{k}\delta \tilde{A}_{k} -ig[A_{k},\delta
\tilde{A}_{k}]$ where $A_{k}(x)=
\theta (x^{-})\frac{i}{g}U(x_{t})\partial _{k}U^{\dagger}(x_{t})$
is the background field. Then the commutator of the background field with
$\delta \tilde{A}_{k}$ will exactly cancel the derivative acting on U's. The
result is that the covariant derivative will reduce to a simple derivative
which acts only on the exponential part of the field.

Following the discussion leading to equation~(\ref{eq:GC}),  we obtain an
implicit expression for the Green's function in Light Cone gauge which we can
write as the sum of the contributions of the type I and type II modes discussed
in the previous section. The former can be written as
\be
G_{ij(I)}^{\alpha\beta ;\alpha^\prime \beta^\prime}(x,y)
&=& \Bigg\{ \delta^{\alpha\gamma}_{ik}
\delta^{\alpha^{\prime}\gamma^{\prime}}_{jl}
-\delta^{\alpha^{\prime}\gamma^{\prime}}_{jl}
 \left(D_{i}\frac{1}{\partial_{-}\partial_{+}}D_{k}\right)
^{\alpha\gamma}_{(x)} -
 \delta^{\alpha\gamma}_{ik}
\left(D_{j}\frac{1}{\partial_{-}
\partial_{+}}D_{l}\right)^{\alpha^{\prime}\gamma^{\prime}}_{(y)}
\nonumber \\
						     &+&
\left(D_{i}\frac{1}{\partial_{-}
\partial_{+}}D_{k}\right)^{\alpha\gamma}_{(x)}
\left(D_{j}\frac{1}{\partial_{-}
\partial_{+}}D_{l}\right)^{\alpha^{\prime}\gamma^{\prime}}_{(y)}\Bigg\}\;\;
\tilde{G}^{\gamma\beta ; \gamma^{\prime}\beta^{\prime}}_{kl(I)}(x,y)  \,,
\label{type1}
\ee
where $\tilde{G}^{\gamma\beta ; \gamma^{\prime}\beta^{\prime}}_{kl(I)}(x,y)$ is
the Green's function in $A^{-}=0$ gauge given by equation~(\ref{OldGreen}). The
subscript $I$ in the above equation denotes the contribution of the type I
class
of small fluctuation fields.

We must add to this expression the contribution from the gauge transform of
$\delta \tilde G^{++}_{(II)}$. This can be expressed as
\be
G_{ij(II)}^{\alpha\beta ;\alpha^\prime \beta^\prime}(x,y)=
\left(D^i{1\over \partial^+}\right)_{(x)}^{\alpha \gamma}\,\,
\left(D^j{1\over \partial^+}\right)_{(y)}^{\alpha \gamma}\,\,
(\delta \tilde G^{++})_{(II)}^{\gamma \beta; \gamma^\prime \beta^\prime}
(x,y) \, ,
\label{type2}
\ee
where
\be
(\delta \tilde G^{++})_{(II)}^{\gamma \beta; \gamma^\prime \beta^\prime} (x,y)=
(\tau _{a})^{\gamma\beta} (\tau _{a})^{\gamma^\prime \beta^\prime}
\int \,{{d^4k}\over {(2\pi)^4}}
{{e^{ik (x-y)}}\over {{(k^-)}^2-i\epsilon}} \, .
\label{green2}
\ee

It is now straightforward to construct similar expressions for non--diagonal
components of the Green's function. Even though only transverse components of
the Light Cone
gauge Green's function are needed to compute the gluon distribution function,
we will give
an implicit expression for all components for the sake of completeness. It is
\be
G^{\mu \nu}(x,y)&=&\Bigg[ \delta ^{\mu \rho}\delta ^{\nu \lambda} + \delta
^{\rho \lambda}
\bigg(D^{\mu}\frac{1}{\partial_- \partial_+}D_{\rho}\bigg)_{(x)}
\bigg(D^{\nu}\frac{1}
{\partial_- \partial_+}D_{\lambda}\bigg)_{(y)} \nonumber \\
& +& \delta ^{\mu \rho}\bigg(\delta ^{\lambda +} - 1\bigg)
\bigg(D^{\nu}\frac{1}{\partial_- \partial_+}D_{\lambda}\bigg)_{(y)}
 \nonumber \\
                &+&\delta ^{\nu \lambda}\bigg(\delta ^{\mu +} - 1\bigg)
\bigg(D^{\mu}\frac{1}
{\partial_- \partial_+}D_{\rho}\bigg)_{(x)} \Bigg]\tilde{G}^{\rho
\lambda}(x,y)\,\,.
\ee
In this expression $\mu , \nu \,=\, 1,2, -\,$ while $\rho \, ,\lambda
\,=\,1,2,-,+ $\,\, and again
$\tilde{G}^{\rho \lambda}(x,y)$ is the Green's function in $A^- =0$ gauge. From
here on, we
will concentrate on the transverse components only.

It is evident from the above equations that in relating the fields in the two
gauges, one has to define $1/\partial^+$ and $1/\partial^-$.  These operators
are
not well defined unless some prescription is adopted for treating the
singularity at $k^+ = 0$ and at $k^- = 0$.  Presumably, there is no physics in
these singularities --when one computes gauge invariant quantities, different
prescriptions should give the same results.  It has however proven most
convenient to choose a Feynman like $i\epsilon $ prescription for regularizing
this singularity, the so called Leibbrandt--Mandelstam prescription~\cite{lm}.
This prescription however spoils the Hermiticity of the field $\delta A^i$.
This is not a new problem since even in the absence of our background
field, such a prescription will destroy the reality properties of $A^-$ in $A^+
= 0$ gauge.  It is indeed unpleasant, but it does make for a Green's function
which is manifestly causal.  The price to be paid for a small lack of
Hermiticity of the fields is small compared to the price of dealing with
non--causal Green's functions, albeit all such ambiguities must come to naught
when computing gauge invariant quantities.

We must now establish our prescriptions for defining the inversion of $\partial
^\pm$.  We first begin with $\frac{1}{k^-}$.  For this op\-er\-a\-tor we use
the
or\-di\-nary Leibbrandt--Mandelstam prescription
\be
	{1 \over k^-} \rightarrow {1 \over {k^- + i\epsilon/k^+}}\, .
\ee
For the operator $1/ \partial^+$, we choose a pre\-scrip\-tion which
guar\-an\-tees that $\delta A^-\mid_{x^- = 0}  = 0$.  This integration choice
is in fact a further specification of the gauge and fixes the residual gauge
freedom completely. This is because if we work in $A^+ = 0$ gauge, we can still
make $x^- $ independent gauge transformations.  We can therefore adjust the
value of $A^-$ at any fixed value of $x^-$ which we choose to be $x^- = 0$. The
virtue of this particular gauge choice is that it makes the source of charge
time independent, since in this case the extended current conservation law is
\be
	\partial^- J^+ = 0\, .
\ee
Our prescription will therefore be
\be
	{1 \over \partial_-} F(x^-) = \int_0^{x^-} dx^{\prime -}
F(x^{\prime-}).
\ee
Let us see what this corresponds to in momentum space.  If we define
\be
	\tilde{F}(k^+) = \int dx^- e^{ik^+x^-} F(x^-)
\ee
then
\be
	\int_0^{x^-}
	dx^{\prime -}F(x^{\prime -}) = \int {{dk^+} \over {2\pi}}
{1 \over {-ik^+}}(e^{-ik^+x^-} -1) \tilde{F} (k^+).
\ee
We see that with this integration prescription, there is never any
singularity in Fourier space as $k^+ \rightarrow 0$.

The reason why this prescription is not often employed is because the types of
fields one ordinarily considers are oscillatory at infinity, and therefore
vanish in the Riemann--Lebesgue sense. In such cases it is more convenient to
choose
$1 /\partial^+ = \int^{x^-}_{-\infty}$ or $1 /\partial^+
= -\int_{x^-}^\infty $
since in the sense described above one can ignore the contribution from
infinity~\cite{soper}. In the Leibbrandt-Mandelstam prescription, positive
frequencies propagate forward in time and negative frequencies propagate
backward in time. For our problem where we have a nucleus acting as a source of
charge, the fields do not vanish at $x^- = \pm \infty$.  There is therefore no
advantage to choosing an integration prescription with one of its integration
limits at either $\pm \infty$.

We can now combine terms together to get an expression for the propagator. The
derivation of $G_{ij(II)}$ from equation~(\ref{type2}) is rather subtle and we
shall elaborate a little on it. We can easily do the $k^{+}$ and $k_{t}$
integrations
in equation~(\ref{green2}) to get
\be
\delta \tilde G^{++}_{(II)} (x,y)= \tau _{a}\tau _{a}\delta ^{(2)}(x_{t}-y_{t})
\delta(x^- -y^-)\int {{dk^-}\over 2\pi} \frac{e^{-ik^-(x^+
-y^+)}}{(k^{-})^{2}-i\epsilon}\,.\nonumber
\ee
It is clear that we have locality in both transverse and minus coordinates.
Hence there will not be any contributions from the regions
$\theta(x^-)\theta(-y^-)$ and $\theta(-x^-)\theta(y^-)$. Then
\be
(D^i)_{(x)}(D^j)_{(y)} \delta \tilde G^{++}_{(II)} &=&
\Bigg[\theta(-x^-)\theta(-y^-)\, \tau_a
\tau_a \,\,\partial_i\partial_j \delta^{(2)}(x_t-y_t)\nonumber \\
&+&\theta(x^-)\theta(y^-)\, F_a(x_t) F_a(y_t)
\,\,\partial_i\partial_j \delta^{(2)}(x_t-y_t)\Bigg] \nonumber \\
&\times & \delta(x^- -y^-)\int {{dk^-}\over 2\pi} \frac{e^{-ik^-(x^+
-y^+)}}{(k^{-})^{2}-i\epsilon}\,.
\ee
We now write the $\delta$--functions in their Fourier representations and
replace
\be
{1 \over {(k^-)^2}} \rightarrow
		     {1 \over k^2} \left(
{{k_t^2} \over {(k^-)^2}} - 2{k^+ \over k^-} \right) \, .
\ee
Finally,  making use of the identity
\be
\int {{dk^+}\over 2\pi} (2k^+)e^{-ik^+(x^- -y^-)} &=& i \Bigg[{\partial\over
\partial x^-}\int {{dk^+}\over 2\pi} e^{-ik^+(x^- -y^-)} \nonumber \\
&-&{\partial\over
\partial y^-}\int {{dk^+}\over 2\pi} e^{-ik^+(x^- -y^-)} \Bigg] \, ,
\ee
we can  write equation~(\ref{type2}) as
\be
G_{ij(II)}^{\alpha\beta ;\alpha^\prime \beta^\prime}(x,y) &=&
\!\!  \int {{d^4p}\over {(2\pi)^4}}
{{e^{ip(x-y)}}\over {p^2-i\epsilon}}
\,\,\bigg[\theta(-x^{-})\theta(-y^{-})\tau_{a}^{\alpha\beta}
\tau_{a}^{\alpha^\prime \beta^\prime} + \theta(x^-)\theta(y^-) \nonumber \\
\!\!\!\!\!&   &\!\!\!\!\!\! F_{a}^{\alpha\beta}(x_{t})
F_{a}^{\alpha^\prime \beta^\prime}(y_{t})
\bigg]\bigg[\frac{p_{i}p_{j}}{p^{-}p^{+}}(e^{ip^{+}x^{-}}\!-1) +
\frac{p_{i}p_{j}}{p^{-}p^{+}}
(e^{-ip^{+}y^{-}}\!-1)\nonumber \\
\!\!\!\!\!&   &\!\!\!\!\!\! + \frac{p_{t}^{2}p_i p_j}{(p^{-}p^{+})^2}
(e^{ip^{+}x^{-}}\!-1)(e^{-ip^{+}y^{-}}\!-1)\bigg] \, .
\label{fntype2}
\ee

The derivation of an analogous expression for $G_{ij(I)}$ in
equation~(\ref{type1}) follows directly from the prescription rules discussed
above. The reader may confirm that each term in equation~(\ref{fntype2})
cancels an identical term in $G_{ij(I)}$. Our final expression for the Green's
function in Light Cone gauge is then
\be
G_{ij}^{\alpha\beta ;\alpha^\prime \beta^\prime}(x,y)
\!\!\!\!\!& = &\! - \int \!{{d^4p}\over {(2\pi)^4}}
{{e^{ip(x-y)}}\over {p^2-i\epsilon}}
\Bigg\{ \delta_{ij} \bigg[\theta(-x^{-})\theta(-y^{-})\tau_{a}^{\alpha\beta}
\tau_{a}^{\alpha^\prime \beta^\prime} + \theta(x^-)\theta(y^-) \nonumber \\
\!\!\!\!\!&   &\!\!\!\!\!\! F_{a}^{\alpha\beta}(x_{t})
F_{a}^{\alpha^\prime \beta^\prime}(y_{t})
\bigg] +  \theta(-x^-) \theta(y^-)
\int \!{{d^2 q_t} \over {(2\pi)^2}} d^2 z_t \,\, e^{i(q^{+}-p^{+})y^{-}}
\nonumber \\
\!\!\!\!\!& \times &\!\!\!\!
e^{i(p_t-q_t)(y_t-z_t)}
 F_{a}^{\alpha \beta}(z_{t})
F_{a}^{\alpha^\prime  \beta^\prime}\!(y_{t})\bigg[\delta_{ij} +
\frac{p_{i}p_{j}}{p^{-}p^{+}}
(e^{ip^{+}x^{-}}\!-1)\nonumber \\
\!\!\!\!\!& + &\!\!\!\!\!\!   \frac{q_{i}q_{j}}{p^{-}q^{+}}
(e^{-iq^{+}y^{-}}\!-1)
+ \frac{p_{i}q_{j}p_{t}\cdot q_{t}}
{(p^{-}p^{+})(p^{-}q^{+})}(e^{ip^{+}x^{-}}\!-1)
(e^{-iq^{+}y^{-}}\!-1)\bigg]\nonumber \\
\!\!\! & + &\!\!\!\!\!
\theta(x^-) \theta(-y^-)\!\!
\int\! {{d^2 q_t} \over {(2\pi)^2}} d^2 z_t
 e^{i(p_t-q_t)(z_t - x_t)} e^{-i(q^{+}-p^{+})x^{-}}\nonumber \\
\!\!\!\!\!&\times &\!\!\!\! F_{a}^{\alpha \beta}(x_{t})
F_{a}^{\alpha^\prime  \beta^\prime}\!(z_{t})\bigg[\delta_{ij} +
\frac{p_{i}p_{j}}{p^{-}p^{+}}(e^{-ip^{+}y^{-}}\!-1)
 +  \frac{q_{i}q_{j}}{p^{-}q^{+}}(e^{iq^{+}x^{-}}\!-1)\nonumber \\
\!\!\!\!\!& + &\!\!\!\!\!\!
\frac{p_{i}q_{j}p_{t}\cdot q_{t}}{(p^{-}p^{+})(p^{-}q^{+})}
(e^{-ip^{+}y^{-}}\!-1)
(e^{iq^{+}x^{-}}\!-1)\bigg]\Bigg\} \, ,
\label{eq:GF}
\ee
where $q^{+}=p^{+}+ \frac{q_{t}^{2}-p_{t}^{2}}{2p^{-}}$\,\,.
Equation~(\ref{eq:GF}) is the main result of this paper. In a forthcoming work,
it will be used to compute the gluon distribution function in the
Weizs\"{a}cker--Williams background field to O($\alpha_s^2$).
There we will show explicitly that when computing the gluon
distribution function from the expression~(\ref{eq:GF}), the only prescription
needed to deal  with poles is the Feynman (causual)~$i\epsilon$ prescription.
In other words, the nature of the additional $p^{-}$ poles (besides the usual
one in $p^{2}-i\epsilon$) in the Green's function is unimportant. Keep in mind
that there are no additional poles in $p^{+}$ since their residues vanish. As a
result, we never have to use any additional $i\epsilon $ \, prescription to
compute the gluon
distribution function to this order. We are not sure if this happens in all
orders of perturbation theory or if it is an amusing coincidence specific to
  O($\alpha_s^2$).

\section{Summary}

In this paper we have computed the propagator for transverse gluons in the
Light Cone gauge $A^+=0$ and in the presence of a background
Weizs\"{a}cker--Williams field produced by the valence quarks of a heavy
nucleus. We have argued that such a propagator can be obtained from the
propagator in the $A^-=0$ gauge by a gauge transformation. This method has the
advantage of avoiding the difficulties introduced by the singular behavior of
the fields at $x^{-}=\pm \infty$ caused by the presence of the sources. The
gauge transformation requires a choice for the inverse of the operators
$\partial^{-}$, $\partial^{+}$. We chose these to be such that the gluon fields
satisfy the boundary condition $\delta A^- |_{x^{-}=0} = 0$ and the usual
Feynman causal behavior. The expression for the gluon propagator in $A^+=0$
gauge is necessary if we want to extract from it the information about the
gluon density which can be compared to experimental measurements.

In the above analysis, there are still many subtle questions remaining about
the independence of physical quantities from the precise choice of the
regularization prescription for the small $k^-$ and $k^+$ singularities. We
have shown that there is a gauge where there are no singularities and that
the subsequent singularities in the Light Cone gauge Green's function
arise from the gauge transformation relations. Then, it is only
sensible that the physical quantities computed here will be independent
of the nature of these singularities.

At the very least it seems we have a consistent prescription. It should be
noted that
because of the gauge choice, there is no ambiguity about the limit $k^+
\rightarrow 0$. The propagator is smooth in this limit as a consequence of the
way we have specified the residual gauge degree of freedom.  If it were only
the
free transverse propagator which was of interest, we would have a non--singular
propagator
at small $k^+$. The problem we have with the distribution function is however
at small $k^-$.
These singularities have been regularized in the only causal way possible, by
an $i \epsilon$
prescription. The question of physical interest is whether these small $k^-$,
or correspondingly, large $k^+$ singularities cause any problems at higher
orders. It has been argued extensively in the literature that the prescription
we chose for $k^-$ is essential to prove the renormalizability of Yang--Mills
theories in Light Cone gauge~\cite{lb2}.

\section*{Acknowledgments}

We acknowledge useful conversations with Mikhail Voloshin. R.V. would like to
thank Mathias Burkhardt for discussions. Two of us, A.A. and J.J.M., would like
to thank R. Rodr\'{\i}guez and R. Madden for their valuable comments. This
research was supported by the U.S. Department of Energy under grants No. DOE
High Energy DE--AC02--83ER40105, No. DOE Nuclear DE--FG02--87ER--40328,
No. DOE Nuclear DE--FG06--90ER--40561, and by the DGAPA/UNAM/M\'exico.

\end{document}